\documentclass[prl,aps,epsfig,superscriptaddress,twocolumn,longbibliography]{revtex4-1}
\usepackage{amsfonts,amssymb,amscd,amsthm}
\usepackage{graphicx}
\usepackage{mathrsfs}
\usepackage[intlimits]{amsmath}
\usepackage[colorlinks, citecolor=red]{hyperref}
\usepackage{etoolbox}
\usepackage{upgreek}
\usepackage{braket}

\begin{document}

\title{Hamiltonian learning for 300 trapped ion qubits with long-range couplings}

\author{S.-A. Guo}
\thanks{These authors contribute equally to this work}%
\affiliation{Center for Quantum Information, Institute for Interdisciplinary Information Sciences, Tsinghua University, Beijing 100084, PR China}

\author{Y.-K. Wu}
\thanks{These authors contribute equally to this work}%
\affiliation{Center for Quantum Information, Institute for Interdisciplinary Information Sciences, Tsinghua University, Beijing 100084, PR China}
\affiliation{Hefei National Laboratory, Hefei 230088, PR China}
\affiliation{Shanghai Qi Zhi Institute, AI Tower, Xuhui District, Shanghai 200232, China}

\author{J. Ye}
\affiliation{Center for Quantum Information, Institute for Interdisciplinary Information Sciences, Tsinghua University, Beijing 100084, PR China}

\author{L. Zhang}
\affiliation{Center for Quantum Information, Institute for Interdisciplinary Information Sciences, Tsinghua University, Beijing 100084, PR China}

\author{Y. Wang}
\affiliation{HYQ Co., Ltd., Beijing 100176, PR China}

\author{W.-Q. Lian}
\affiliation{HYQ Co., Ltd., Beijing 100176, PR China}

\author{R. Yao}
\affiliation{HYQ Co., Ltd., Beijing 100176, PR China}

\author{Y.-L. Xu}
\affiliation{Center for Quantum Information, Institute for Interdisciplinary Information Sciences, Tsinghua University, Beijing 100084, PR China}

\author{C. Zhang}
\affiliation{HYQ Co., Ltd., Beijing 100176, PR China}

\author{Y.-Z. Xu}
\affiliation{Center for Quantum Information, Institute for Interdisciplinary Information Sciences, Tsinghua University, Beijing 100084, PR China}

\author{B.-X. Qi}
\affiliation{Center for Quantum Information, Institute for Interdisciplinary Information Sciences, Tsinghua University, Beijing 100084, PR China}

\author{P.-Y. Hou}
\affiliation{Center for Quantum Information, Institute for Interdisciplinary Information Sciences, Tsinghua University, Beijing 100084, PR China}
\affiliation{Hefei National Laboratory, Hefei 230088, PR China}

\author{L. He}
\affiliation{Center for Quantum Information, Institute for Interdisciplinary Information Sciences, Tsinghua University, Beijing 100084, PR China}
\affiliation{Hefei National Laboratory, Hefei 230088, PR China}

\author{Z.-C. Zhou}
\affiliation{Center for Quantum Information, Institute for Interdisciplinary Information Sciences, Tsinghua University, Beijing 100084, PR China}
\affiliation{Hefei National Laboratory, Hefei 230088, PR China}

\author{L.-M. Duan}
\email{To whom correspondence should be addressed; E-mail: lmduan@tsinghua.edu.cn.}
\affiliation{Center for Quantum Information, Institute for Interdisciplinary Information Sciences, Tsinghua University, Beijing 100084, PR China}
\affiliation{Hefei National Laboratory, Hefei 230088, PR China}
\affiliation{New Cornerstone Science Laboratory, Beijing 100084, PR China}

\begin{abstract}
Quantum simulators with hundreds of qubits and engineerable Hamiltonians have the potential to explore quantum many-body models that are intractable for classical computers. However, learning the simulated Hamiltonian, a prerequisite for any applications of a quantum simulator, remains an outstanding challenge due to the fast increasing time cost with the qubit number and the lack of high-fidelity universal gate operations in the noisy intermediate-scale quantum era. Here we demonstrate the Hamiltonian learning of a two-dimensional ion trap quantum simulator with $300$ qubits. We employ global manipulations and single-qubit-resolved state detection to efficiently learn the all-to-all-coupled Ising model Hamiltonian, with the required quantum resources scaling at most linearly with the qubit number. Our work paves the way for wide applications of large-scale ion trap quantum simulators.
\end{abstract}

\maketitle

\section{Introduction}
Quantum computers and quantum simulators have reached the stage of coherently manipulating hundreds of qubits \cite{kim2023evidence,Bluvstein2024,bohnet2016quantum,guo2023siteresolved}, and quantum advantage over classical computers has been demonstrated on random sampling tasks \cite{arute2019quantum,jiuzhang2020,PhysRevLett.127.180501,madsen2022quantum}. To achieve the next milestone of quantum advantage with practical utility, quantum simulation of many-body dynamics is one of the most promising candidates \cite{Cirac2012,RevModPhys.86.153,RevModPhys.93.025001,RevModPhys.94.015004}. However, verifying the quantum simulation results for such classically intractable problems is a notoriously challenging task. To completely characterize the simulated dynamics and to check if it follows the desired Hamiltonian evolution, a quantum process tomography requires a time cost that grows exponentially with the system size $N$ \cite{nielsen2000quantum}. To reduce this complexity, various Hamiltonian learning algorithms have been developed \cite{PhysRevLett.107.210404,PhysRevLett.112.190501,Wang_2015,Wang2017,Krastanov_2019,hou2019cpl,Qi2019determininglocal,PhysRevLett.122.020504,evans2019scalable,Bairey_2020,Gentile2021,PhysRevA.105.023302,PRXQuantum.3.030345,Yu2023robustefficient,Joshi2023,Haah2024,bakshi2024structure}, utilizing different types of a priori knowledge about the system like locality \cite{PhysRevLett.107.210404,Wang_2015,PhysRevA.105.023302,hou2019cpl,Qi2019determininglocal,PhysRevLett.122.020504,evans2019scalable,Bairey_2020,Joshi2023,Haah2024,bakshi2024structure} and sparsity \cite{Krastanov_2019,Gentile2021,Yu2023robustefficient}, or with the help of certain steady states or thermal states \cite{Qi2019determininglocal,PhysRevLett.122.020504,evans2019scalable,Bairey_2020,Haah2024} or other trusted quantum devices \cite{PhysRevLett.112.190501,Wang2017,PRXQuantum.3.030345} as quantum resources. Nevertheless, a $\mathrm{poly}(N)$ time complexity is generally inevitable, which becomes a considerable cost for the large-scale quantum simulators. Besides, many of the learning algorithms requires individually addressed quantum gates \cite{PhysRevLett.107.210404,Wang_2015,Qi2019determininglocal,PhysRevLett.122.020504,evans2019scalable,Bairey_2020,PRXQuantum.3.030345,Yu2023robustefficient,Joshi2023,Haah2024,bakshi2024structure}, which may not be available on the noisy intermediate-scale quantum (NISQ) devices.

On the other hand, quantum simulators based on arrays of atoms \cite{Saffman_2016,Wu_2021,ebadi2021quantum} or ions \cite{RevModPhys.93.025001,guo2023siteresolved} can naturally support single-shot readout of all the $N$ qubits, giving $N$ bits of information per trial. This can largely mitigate the required quantum simulation resources by up to a factor of $N$. Following this idea, the coherent imaging spectroscopy technique has been developed where all the $O(N^2)$ coefficients in a fully connected Ising model can be determined from $O(N)$ frequency scans with global quantum manipulation and individual state detection \cite{CIS}. However, this scheme is subjected to lower signal-to-noise ratio as the state preparation error accumulates with increasing $N$. Also the time for each frequency scan may need to scale polynomially with $N$ in order to resolve the decreasing energy gaps. Due to these restrictions, this scheme has only been applied to measure an 8-spin Hamiltonian \cite{CIS} and to partially verify the theoretical calculations for $61$ qubits \cite{PRXQuantum.4.010302}.

Here, we report the Hamiltonian learning of an ion trap quantum simulator with $300$ qubits. We use global laser and microwave operations to perform a Ramsey-type experiment with various evolution times under the desired Ising Hamiltonian. We extract single-spin and two-spin observables from the single-shot measurements to fit all the $O(N^2)$ ($44850$ in total) Ising coupling coefficients, and we test the learning results on independent data to show that there is no significant overfitting. We further compare this general model to a physically guided one with $O(N)$ parameters. By independently calibrating the collective phonon modes of the ions and the laser intensity, the latter learning algorithm can achieve a similar test error as the former, with an improved scaling for the required sample size. We further compute the dynamics of higher-order spin correlations from the learned Hamiltonian and validate the learning results from their consistency with the experimental data. Our method can be applied to even larger ion crystals and paves the way for the applications of the ion trap quantum simulators on various NISQ algorithms \cite{RevModPhys.86.153,RevModPhys.94.015004}.

\section{Experimental scheme}
\begin{figure*}
	\centering
	\includegraphics[width=\linewidth]{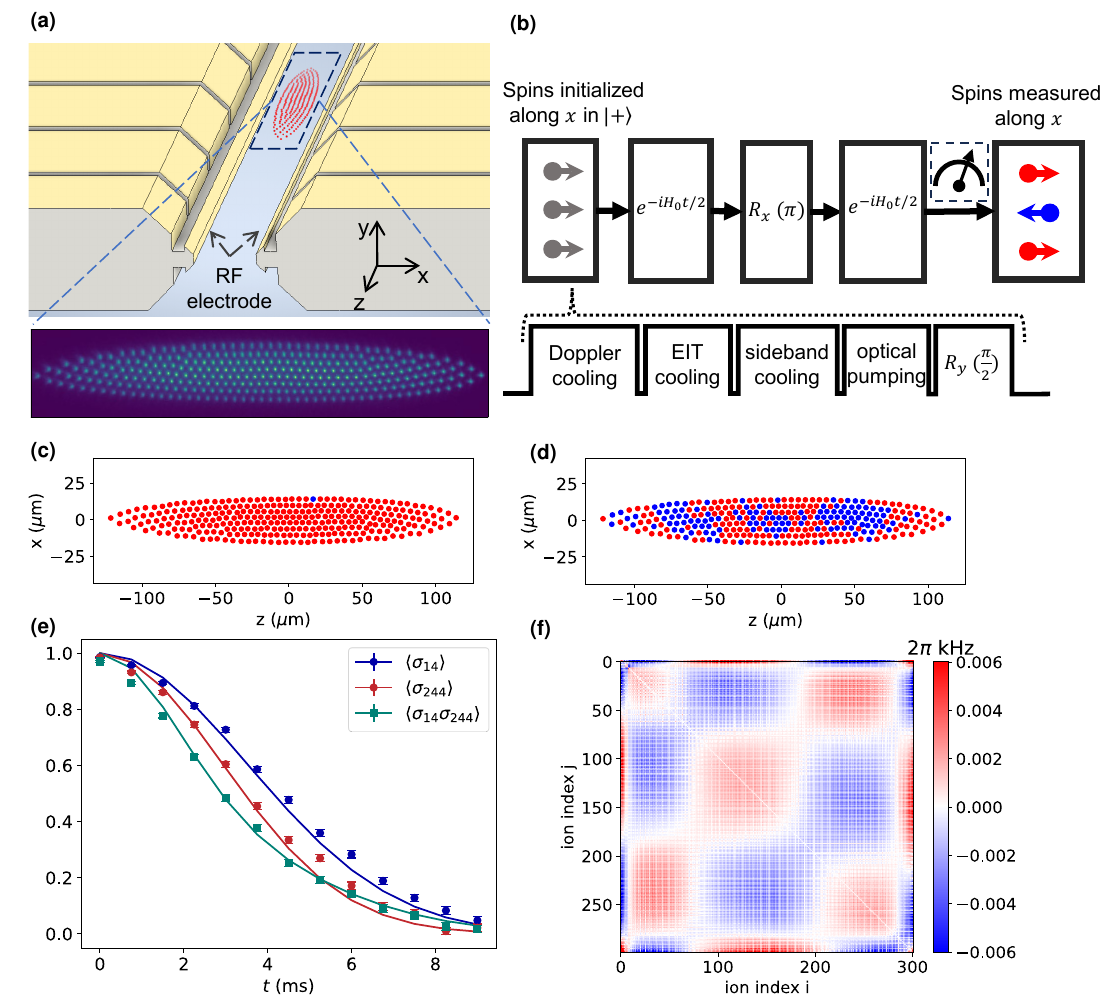}
	\caption { \textbf{Experimental scheme.} (a) We use a cryogenic monolithic ion trap to obtain a 2D crystal of $N=300$ $^{171}\mathrm{Yb}^+$ ions. (b) We perform a Ramsey-type experiment to learn the long-range Ising Hamiltonian using global control and individual readout of the qubits. We initialize all the ions in $|0\rangle$ by laser cooling and optical pumping, and rotate them into $|+\rangle$ by a microwave $\pi/2$ pulse. The system is then evolved under the Ising Hamiltonian $H_0$ for various times $t$ with a spin echo in the middle to cancel the longitudinal fields. Finally we measure all the ions in the $\sigma_x$ basis via another microwave $\pi/2$ pulse followed by the state-dependent fluorescence detection. (c) Typical single-shot measurement result at $t=0\,$ms. (d) Typical single-shot measurement result at $t=9\,$ms. (e) We use all the single-spin (``magnetization'') and two-spin (``correlation'') observables to fit the Ising coupling coefficients. Typical fitting results (solid curves) are compared with the measured data (dots for magnetization and squares for correlation with one standard deviation error bar) for an arbitrarily chosen ion pair. (f) The fitted Ising coupling coefficients $J_{ij}$ when the laser couples dominantly to the fifth highest phonon mode.} \label{fig:1}
\end{figure*}

Our experimental setup is sketched in Fig.~\ref{fig:1}(a) with a 2D crystal of 300 ${}^{171}\mathrm{Yb}^+$ ions in a cryogenic monolithic ion trap \cite{guo2023siteresolved}. The qubits are encoded in the hyperfine ground states $|0\rangle\equiv |S_{1/2},F=0,m_F=0\rangle$ and $|1\rangle\equiv |S_{1/2},F=1,m_F=0\rangle$. By applying counter-propagating $411\,$nm global laser beams on the ions perpendicular to the 2D crystal, we can generate a long-range Ising Hamiltonian $H_0=\sum_{i<j}J_{ij}\sigma_z^i \sigma_z^j$ intermediated by the transverse (drumhead) phonon modes \cite{guo2023siteresolved,PhysRevA.103.012603}. When supplemented by a global microwave resonant to the qubit frequency, a transverse-field Ising model $H=H_0+B \sum_{i}\sigma_x^i $ can be obtained \cite{guo2023siteresolved} with wide applications in quantum many-body physics \cite{RevModPhys.93.025001} and NISQ algorithms \cite{RevModPhys.86.153,RevModPhys.94.015004}. Since the transverse field $B$ can be accurately controlled and separately calibrated in experiments, here we focus on the calibration of the Ising coupling coefficients $J_{ij}$'s, which represents the challenging part of the Hamiltonian $H$ to be learned. Theoretically, with the 2D crystal locating on an equiphase surface of the laser (see Supplementary Materials) and under the virtual excitation condition of the phonon modes \cite{guo2023siteresolved,RevModPhys.93.025001}, the coupling coefficients can be given as
\begin{equation}
J_{ij} = \sum_k \frac{1}{8(\mu-\omega_k)} \eta_k^2 b_{ik} b_{jk} \Omega_i \Omega_j, \label{eq:1}
\end{equation}
where $\mu$ is the laser detuning, $\omega_k$ the frequency of the $k$-th mode, $\eta_k$ the Lamb-Dicke parameter, $\Omega_i$ the laser-induced AC Stark shift on the $i$-th ion, and $b_{ik}$ the normalized mode vector.

The experimental sequence for Hamiltonian learning is shown in Fig.~\ref{fig:1}(b) where we perform a Ramsey-type experiment to extract information about $H_0$ by tuning the transverse $B$ field to zero. We initialize the spins in $|+\rangle$ by a global microwave $\pi/2$ pulse, evolve them under $H_0$ for time $t$, and finally measure all the spins in the $\sigma_x$ basis by another microwave $\pi/2$ pulse. To remove the influence of possible longitudinal fields $H^\prime=\sum_i h_i \sigma_z^i$ (which can be calibrated separately if needed), we apply a $\pi$ pulse in the middle which commutes with the desired $H_0$. A typical single-shot measurement result at $t=0$ is shown in Fig.~\ref{fig:1}(c) with a few random spins being flipped due to the about $0.7\%$ state-preparation-and-measurement (SPAM) errors using the electron shelving technique \cite{Roman2020,edmunds2020scalable,yang2022realizing}. Similarly, in Fig.~\ref{fig:1}(d) we show a typical single-shot measurement result at $t=9\,$ms when the Ising interaction is dominated by the pattern of the fifth highest phonon mode.
By further averaging over $M$ experimental trials, we can estimate any $k$-body spin correlation functions. Here we focus on the single-spin (``magnetization'') and two-spin (``correlation'') observables and use them to fit all the $J_{ij}$'s, as shown in Fig.~\ref{fig:1}(e) and Fig.~\ref{fig:1}(f). Later we will use higher-order correlations to verify the Hamiltonian learning results.

Suppose we take data from $T$ different evolution times. The $NT$ magnetizations and $N(N-1)T/2$ correlations contain sufficient information to learn the $N(N-1)/2$ Ising coupling coefficients. Actually, in principle even the early-time dynamics $\langle \sigma_i (t) \sigma_j(t)\rangle-\langle\sigma_i(t)\rangle\langle\sigma_j(t)\rangle\approx 4|J_{ij}|^2 t^2$ is sufficient to determine the magnitude of all the $J_{ij}$'s. Here we use longer evolution time and the analytical formulae for the magnetizations and correlations (see Supplementary Materials) to fit the Ising coefficients so that the results will be more robust to the experimental noises. Due to the symmetry of the Hamiltonian and the initial state, the measured dynamics will be invariant if we change $J_{ij} \to -J_{ij}$ ($\forall j \ne i$) for any given spin $i$. In other words, there exist an exponential number of equivalent solutions under our experimental sequence, which in principle can be distinguished by preparing different initial states. However, note that these equivalent patterns are discrete and well-separated from each other given that each spin is strongly coupled to at least one of other spins. Therefore we can simply use the theoretical predictions like Eq.~(\ref{eq:1}) as the starting point of the fitting to break their symmetry.

\begin{figure*}
	\centering
	\includegraphics[width=\linewidth]{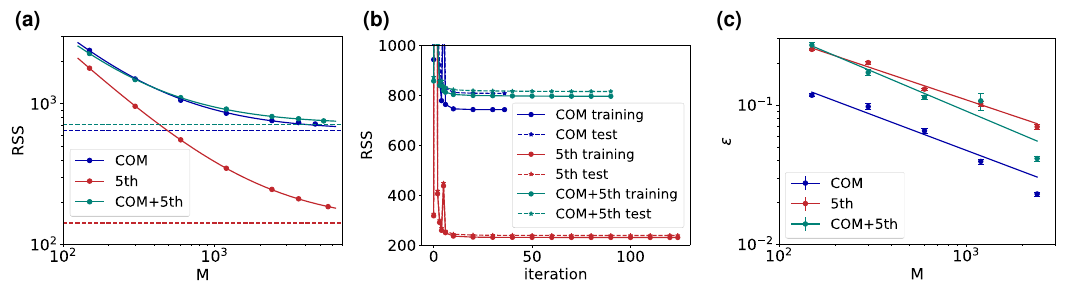}
	\caption {\textbf{Hamiltonian learning with $O(N^2)$ parameters.} (a) Residual sum of squares (RSS) for $N=300$ ions and $T=13$ time steps versus the sample size $M$ at each step. Whether the Ising Hamiltonian comes from coupling to the first (blue), fifth (red) or both (green) highest phonon modes, the experimental results can be well fitted by $y=ax^{-1}+b$ (solid curves) where $b>0$ (horizontal dashed lines) corresponds to the deviation from the theoretical model due to experimental imperfections. After discarding the data when the configuration of the ion crystal is changed during the experimental sequence (see Supplementary Materials), we get in total $M=4638,\,5640,\,5272$ samples for the three Hamiltonians, respectively. (b) Typical learning curves when we randomly divide the samples at each time point into two equal halves as the training and the test sets. We fit the Ising coupling coefficients $J_{ij}$'s using the training set by iteratively minimizing their RSS (training-RSS). For each step of iteration, we record the $J_{ij}$'s and compute the RSS on the test set (test-RSS). The training-RSS and the test-RSS show the same tendency and comparable final values, indicating there is no significant overfitting in the obtained $J_{ij}$'s. (c) Precision $\epsilon$ of the learning results versus the sample size $M$. We take two disjoint sets of data, each with $M$ samples, to learn the Ising coupling coefficients $J_{ij}^{(1)}$ and $J_{ij}^{(2)}$. Then we compute the relative energy difference between the two learned Hamiltonians for 1000 randomly sampled spin configurations. Finally we further average over five random choices of the disjoint data sets, and we use the standard deviation as the error bars. A scaling of $\epsilon \propto M^{-\alpha}$ is fitted with $\alpha \in [0.45, 0.57]$, close to the theoretical scaling of $M^{-0.5}$.} \label{fig:2}
\end{figure*}

We demonstrate this Hamiltonian learning algorithm for different Ising coupling coefficients in Fig.~\ref{fig:2}, with the laser dominantly coupled to the highest (center-of-mass) or the fifth phonon mode, or using two frequency components to couple to both of them \cite{guo2023siteresolved}. As shown in Fig.~\ref{fig:2}(a), the residual sum of squares (RSS) of the least square fitting follows a $1/M$ scaling (solid curves) versus the sample size $M$ at each time point due to the statistical fluctuation. There also exists a nonzero RSS in the limit $M\to \infty$ due to the deviation from the ideal fitting model like the SPAM error, the nonzero phonon excitation and the spin dephasing due to high-frequency noises. (The dominant dephasing source of a shot-to-shot laser intensity fluctuation has been included in the theoretical model as described in Supplementary Materials.) If we view the least square fitting as an optimization problem, the RSS is already close to the $M\to \infty$ case (horizontal dashed lines) with about $M=5000$ samples. However, this does not exclude the possible overfitting in the obtained Ising coefficients. For this purpose, we further plot the learning curve during the training process in Fig.~\ref{fig:2}(b). We randomly split the data at each time point into two equal halves as the training and the test sets. We minimize the RSS of the training set by iterative algorithms and also compute the corresponding RSS on the test set at each step. As we can see, for all the three Hamiltonians to be learned, the RSS for the training and the test sets show the same tendency and similar final values. This suggests that our sample size is large enough to avoid overfitting in the learning results.

Nevertheless, it is still desirable to have more samples to improve the precision. To quantify the precision of the learned Ising Hamiltonian, here we define a relative energy difference $\epsilon(J^{(1)},J^{(2)})\equiv \langle |E(J^{(1)})-E(J^{(2)})|\rangle/\sqrt{\delta E(J^{(1)}) \cdot \delta E(J^{(2)})}$, where the numerator is the energy difference between two sets of Ising coefficients $J_{ij}^{(1)}$'s and $J_{ij}^{(2)}$'s averaged over all the spin configurations, and the denominator consists of the standard deviation of the energy for the two sets, again over all the spin configurations (see Supplementary Materials for details). Roughly speaking, this quantity characterizes the phase difference accumulated during the typical timescale of the two Hamiltonians and has a scaling of $\sqrt{N}$ with the system size. Now we can take two disjoint sets of data, each with $M$ samples, to learn the Ising coupling coefficients $J^{(1)}$ and $J^{(2)}$, and compute their relative energy difference. We further average over random choices of the data sets to quantify the precision and fit a scaling $\epsilon\propto 1/\sqrt{M}$ as shown in Fig.~\ref{fig:2}(c). From the fitting results, we can estimate that about $10^4$ samples at each time point will be needed to reach a precision of $1\%$ for the $N=300$ qubits. Note that the above definition of the precision is for general quantum dynamics and average spin configurations. In many cases we will be interested in the ground states, then the undesired scaling of $\sqrt{N}$ can be removed and the precision can be largely improved (see Supplementary Materials). However, for general Ising Hamiltonian the ground state energy may be difficult to evaluate, therefore here we still use the above definition of the precision, while recognizing that the actual precision may be better for certain tasks.

\begin{figure*}
	\centering
	\includegraphics[width=\linewidth]{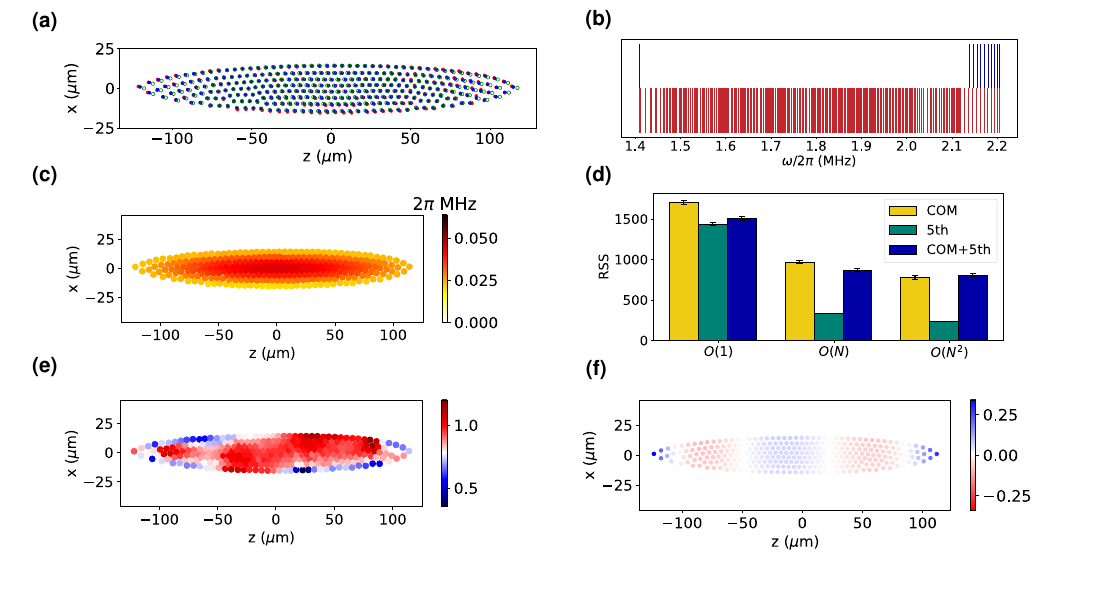}
	\caption {\textbf{Hamiltonian learning with $O(N)$ physically guided parameters.} (a) and (b) Calibration of anharmonic trap potential. (a) Blue dots are the measured ion positions from the CMOS camera. Green hollow circles are the best fit assuming a quadratic potential in the $xz$ plane. Red dots further include cubic and quartic terms in the potential. (b) Blue vertical lines are the measured transverse phonon mode frequencies. The phonon spectrum becomes too dense to resolve at the low-frequency side, so we measure the 10 highest phonon mode frequencies and one lowest frequency. Red vertical lines are the best-fitted theoretical mode frequencies with cubic and quartic potentials. (c) AC Stark shift $\Omega_i$ of individual ions from experimentally calibrated Rabi rates. They can further be fitted by a Gaussian profile with a full width at half maximum of $241\,\mu$m along the $z$ direction and $29\,\mu$m along the $x$ direction. (d) Comparison of test-RSS for Hamiltonian learning with $O(1)$, $O(N)$ and $O(N^2)$ parameters. Here $O(N^2)$-scheme is the method in Fig.~\ref{fig:2}, $O(1)$-scheme is to compute the $J_{ij}$ coefficients from the above calibrated phonon modes and laser intensities, while $O(N)$-scheme is to use the calibrated phonon modes but fit $\Omega_i$'s as free parameters. The yellow, green and blue bars correspond to the Ising Hamiltonian when coupling to the first, fifth or both phonon modes, respectively. Error bars represent one standard deviation when randomly splitting the data into training and test sets for 10 times. (e) Ratio of the fitted $\Omega_i$'s for the $O(N)$-scheme to the calibrated values in the $O(1)$-scheme, when the laser couples mainly to the fifth phonon mode. (f) Mode structure $b_{ik}$ of the fifth highest phonon mode.} \label{fig:3}
\end{figure*}

The above scaling of $\epsilon\propto \sqrt{N/M}$ suggests that to reach the desired precision for large-scale quantum simulators, the required sample size $M$ may scale linearly with the qubit number $N$. This can also be understood as follows: Although we use $O(N^2)$ observables (including magnetizations and correlations) to learn the $O(N^2)$ parameters of the Hamiltonian, note that they are computed from $M$ single-shot measurement results which contain at most $NM$ bits of information. In this sense, $M\sim O(N)$ samples will be necessary to estimate all the $O(N^2)$ parameters. To overcome this general scaling, additional knowledge about the physical system must be exploited to parameterize the Hamiltonian more economically. According to Eq.~(\ref{eq:1}), one possibility is to calibrate the phonon modes and the laser intensities on all the ions. As shown in Fig.~\ref{fig:3}(a) and (b) with more details in Supplementary Materials, we can fit the anharmonic trap potential up to the fourth order from the measured equilibrium positions of the ions and a few phonon modes that can be resolved, and further compute all the phonon mode structures theoretically. Also, in Fig.~\ref{fig:3}(c) we calibrate the laser intensity on individual ions by driving their carrier Rabi oscillations between the $S_{1/2}$ and $D_{5/2}$ levels. However, if we directly compute the Ising coupling coefficients using Eq.~(\ref{eq:1}) (or add up two sets of such computed coefficients when applying two frequency components), the performance of the learned Hamiltonian is typically not satisfactory, with much higher RSS on the test set as shown by the $O(1)$-scheme in Fig.~\ref{fig:3}(d).

To understand the deviation from the above $O(N^2)$-scheme, we fix the calibrated phonon modes and turn the laser-induced AC Stark shift $\Omega_i$ into $N$ fitting parameters. From Fig.~\ref{fig:3}(d) we can see that with these $O(N)$ fitting parameters, the RSS on the test data already becomes close to the $O(N^2)$-scheme. We plot the ratio between the fitted $\Omega_i$ and the measured ones in Fig.~\ref{fig:3}(e) when the laser couples dominantly to the fifth phonon mode. The discrepancy is most significant near the nodes of the fifth mode as shown in Fig.~\ref{fig:3}(f) where $b_{ik}\approx 0$, and near the edge of the 2D crystal with large micromotion of the ions. This suggests that the deviation between the $O(1)$-scheme and the $O(N^2)$-scheme is still restricted by the inaccurate calibration of the phonon modes, and may be improved in the future by including the micromotion into the theoretical model. It also means that the $O(N)$ fitting parameters do not have the physical meaning of the laser intensity, but are to compensate the miscalibrated phonon modes. Therefore, for the Hamiltonian when the laser has two frequency components to couple to two phonon modes, we should introduce $N$ independent fitting parameters for each mode.

\section{Predicting high-order correlations}

\begin{figure*}
	\centering
	\includegraphics[width=\linewidth]{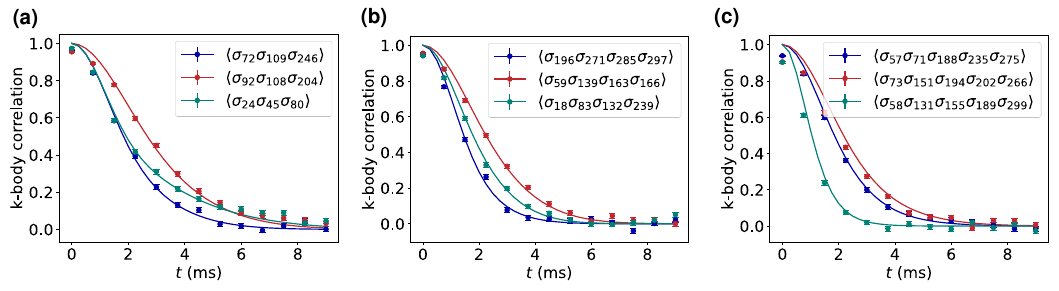}
	\caption {\textbf{Validate Hamiltonian learning results by higher-order correlations.} We further use the Hamiltonian, learned from single-spin and two-spin observables, to predict $k$-body spin correlations with (a) $k=3$, (b) $k=4$ and (c) $k=5$. We compare the theoretical predictions (solid curves) with experimental results (dots) for three randomly chosen sets of ion indices. Each experimental point is averaged over $M=5640$ samples. Here we use the Hamiltonian when the laser couples dominantly to the fifth highest phonon mode as an example.} \label{fig:4}
\end{figure*}

Apart from explaining all the magnetizations and the correlations, the learned Hamiltonian should also be able to predict the dynamics of higher-order spin correlations. As shown in Supplementary Materials, for our experimental sequence, each $k$-body correlation can be computed analytically with a time cost of $O(2^k)$, and there are $O(N^k)$ such terms for small $k$. Therefore, instead of testing all of them or even including them in the Hamiltonian learning process, here we choose to validate the learning results using a few randomly selected sets of ions. As shown in Fig.~\ref{fig:4}, we use the learning results of the $O(N^2)$-scheme, which come from all the single-spin and two-spin observables, to predict the dynamics of $k$-body correlations with $k=3,\,4,\,5$. For the arbitrarily chosen ion indices, we see good agreement between the theoretical and experimental results, which again shows the validity and the insignificant overfitting in the learning results.

\section{Discussion}
To sum up, in this work we demonstrate the Hamiltonian learning of a long-range Ising model with $N=300$ spins on a 2D ion trap quantum simulator. By exploiting global quantum operations and single-qubit-resolved measurements, we can learn a general Ising model with $O(N^2)$ parameters and the required sample size scales at most linearly with the qubit number. By independently calibrating the phonon modes of the ion crystal, a more efficient scheme with $O(N)$ parameters can achieve a similar value of the test error, although the fitting parameters may lack a good physical interpretation and may be related to the miscalibration of the phonon modes. In the future by including higher-order nonlinear potential and the micromotion into the description of the phonon modes, we may improve the accuracy of the $O(1)$-scheme to be comparable to the $O(N^2)$-scheme, thus further enhance the efficiency of the Hamiltonian learning task.

Our method can also be extended to contain a site-dependent longitudinal field $H^\prime=\sum_i h_i \sigma_z^i$, which is required in general applications like the formulation of a quadratic unconstrained binary optimization problem \cite{Lucas2014Ising}. Note that such a longitudinal field is cancelled by the spin echo in our experimental sequence. Therefore, a straightforward method is to first use the sequence with the spin echo to learn all the Ising coupling coefficients $J_{ij}$'s, and then fix them and execute another set of sequences without the spin echo to learn the additional $N$ parameters $h_i$'s, following the general analytical formulae in Supplementary Materials.

\begin{acknowledgments}
This work was supported by Innovation Program for Quantum Science and Technology (2021ZD0301601, 2021ZD0301605), the Shanghai Qi Zhi Institute, Tsinghua University Initiative Scientific Research Program, and the Ministry of Education of China. L.M.D. acknowledges in addition support from the New Cornerstone Science Foundation through the New Cornerstone Investigator Program. Y.K.W. acknowledges in addition support from Tsinghua University Dushi program. C.Z. acknowledges support from the Shui-Mu Scholar postdoctoral fellowship from Tsinghua University. P.Y.H. acknowledges the start-up fund from Tsinghua University.
\end{acknowledgments}

%

\end{document}


\title{Supplementary Materials for ``Hamiltonian learning for 300 trapped ion qubits with long-range couplings''}

\author{S.-A. Guo}
\thanks{These authors contribute equally to this work}%
\affiliation{Center for Quantum Information, Institute for Interdisciplinary Information Sciences, Tsinghua University, Beijing 100084, PR China}

\author{Y.-K. Wu}
\thanks{These authors contribute equally to this work}%
\affiliation{Center for Quantum Information, Institute for Interdisciplinary Information Sciences, Tsinghua University, Beijing 100084, PR China}
\affiliation{Hefei National Laboratory, Hefei 230088, PR China}
\affiliation{Shanghai Qi Zhi Institute, AI Tower, Xuhui District, Shanghai 200232, China}

\author{J. Ye}
\affiliation{Center for Quantum Information, Institute for Interdisciplinary Information Sciences, Tsinghua University, Beijing 100084, PR China}

\author{L. Zhang}
\affiliation{Center for Quantum Information, Institute for Interdisciplinary Information Sciences, Tsinghua University, Beijing 100084, PR China}

\author{Y. Wang}
\affiliation{HYQ Co., Ltd., Beijing 100176, PR China}

\author{W.-Q. Lian}
\affiliation{HYQ Co., Ltd., Beijing 100176, PR China}

\author{R. Yao}
\affiliation{HYQ Co., Ltd., Beijing 100176, PR China}

\author{Y.-L. Xu}
\affiliation{Center for Quantum Information, Institute for Interdisciplinary Information Sciences, Tsinghua University, Beijing 100084, PR China}

\author{C. Zhang}
\affiliation{HYQ Co., Ltd., Beijing 100176, PR China}

\author{Y.-Z. Xu}
\affiliation{Center for Quantum Information, Institute for Interdisciplinary Information Sciences, Tsinghua University, Beijing 100084, PR China}

\author{B.-X. Qi}
\affiliation{Center for Quantum Information, Institute for Interdisciplinary Information Sciences, Tsinghua University, Beijing 100084, PR China}

\author{P.-Y. Hou}
\affiliation{Center for Quantum Information, Institute for Interdisciplinary Information Sciences, Tsinghua University, Beijing 100084, PR China}
\affiliation{Hefei National Laboratory, Hefei 230088, PR China}

\author{L. He}
\affiliation{Center for Quantum Information, Institute for Interdisciplinary Information Sciences, Tsinghua University, Beijing 100084, PR China}
\affiliation{Hefei National Laboratory, Hefei 230088, PR China}

\author{Z.-C. Zhou}
\affiliation{Center for Quantum Information, Institute for Interdisciplinary Information Sciences, Tsinghua University, Beijing 100084, PR China}
\affiliation{Hefei National Laboratory, Hefei 230088, PR China}

\author{L.-M. Duan}
\email{To whom correspondence should be addressed; E-mail: lmduan@tsinghua.edu.cn.}
\affiliation{Center for Quantum Information, Institute for Interdisciplinary Information Sciences, Tsinghua University, Beijing 100084, PR China}
\affiliation{Hefei National Laboratory, Hefei 230088, PR China}
\affiliation{New Cornerstone Science Laboratory, Beijing 100084, PR China}

\maketitle

\section{Experimental setup}
We use a monolithic 3D Paul trap \cite{guo2023siteresolved,Brownnutt_2006,https://doi.org/10.1002/qute.202000068} with an RF frequency of $\omega_{\mathrm{rf}}=2\pi\times 35.280\,$MHz at a cryogenic temperature of $6.1\,$K. To obtain a 2D crystal of $N=300$ ${}^{171}\mathrm{Yb}^+$ ions, we use trap frequencies of $(\omega_x,\omega_y,\omega_z)=2\pi\times(0.623,2.20,0.147)\,$MHz where the $z$ direction is the axial direction without micromotion as shown in Fig.~1(a) of the main text.

All the laser beams propagate in the micromotion-free directions in the $yz$ plane to be insensitive to the inevitable micromotion of the 2D crystal along the $x$ direction. We use a global $370\,$nm laser beam for Doppler cooling, optical pumping and qubit state detection by turning on or off $14.7\,$GHz and $2.1\,$GHz electro-optic modulators (EOMs).
Another two $370\,$nm laser beams are used for EIT cooling with $\pi$ and $\sigma^+$ polarizations perpendicular to each other \cite{PhysRevLett.125.053001}. They have a blue detuning of about $86\,$MHz from the transition between $|S_{1/2},F=1,m_F=0\rangle$ ($|S_{1/2},F=1,m_F=-1\rangle$) and $|P_{1/2},F=0,m_F=0\rangle$. We further use a global $411\,$nm laser beam with a linewidth of about $1\,$kHz, perpendicular to the ion crystal, for the sideband cooling of the transverse phonon modes.

The imaging system, with an NA of 0.33, is also perpendicular to the 2D ion crystal. We use a CMOS camera to collect the fluorescence from individual ions. We use electron shelving for the single-shot state detection by first converting the population in the $|S_{1/2},F=0,m_F=0\rangle$ state to the $D_{5/2}$ and $F_{7/2}$ levels through global $411\,$nm and $3432\,$nm laser \cite{edmunds2020scalable,yang2022realizing}. Then we count the fluorescence photon from individual ions under the global $370\,$nm laser with an exposure time of $1.5\,$ms. The state-preparation-and-measurement infidelity is about $0.7\%$, mainly due to the imperfect shelving under inhomogeneous laser beams.

\begin{figure}
    \centering
    \includegraphics[width=\linewidth]{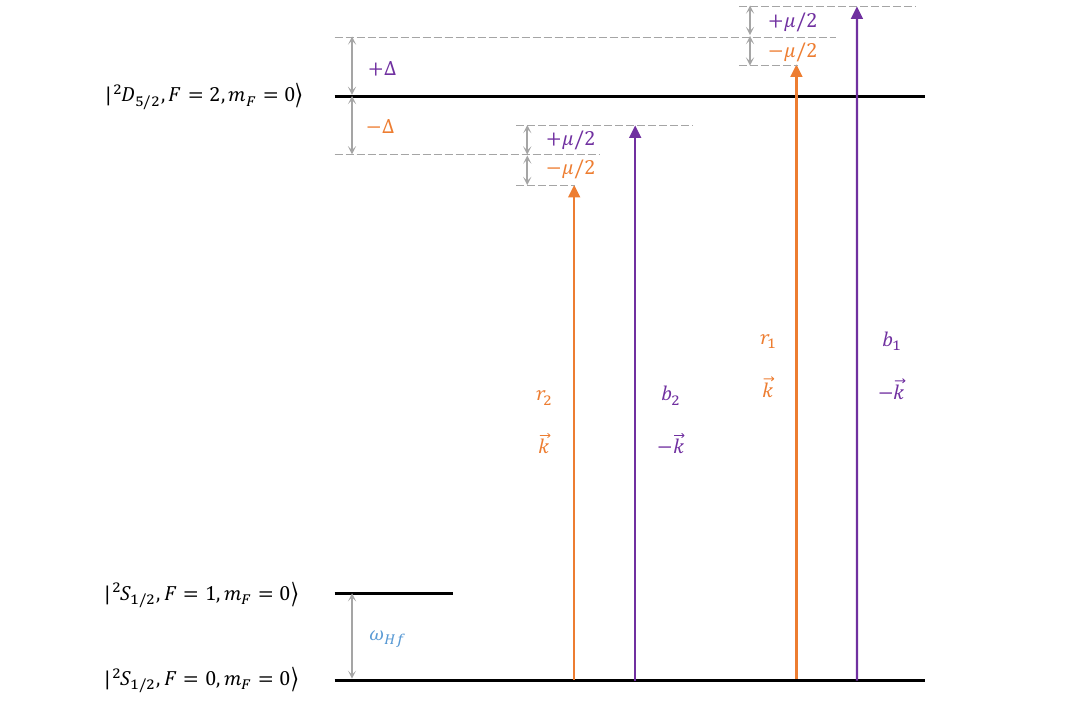}
    \caption{\label{fig:Ising} Laser configuration to generate the long-range Ising model Hamiltonian.}
\end{figure}

We use two counter-propagating $411\,$nm laser beams to generate the long-range Ising interaction. As shown in Fig.~\ref{fig:Ising}, each laser beam has two frequency components on the two sides of the $|S_{1/2},F=0,m_F=0\rangle$ to $|D_{5/2},F=2,m_F=0\rangle$ transition, such that their time-independent AC Stark shift can be roughly cancelled. On the other hand, the beat note of these two beams creates an AC Stark shift varying in time and space, which can create a spin-dependent force \cite{leibfried2003experimental,Lee_2005} and further an Ising-type spin-spin interaction when the phonon modes are adiabatically eliminated \cite{RevModPhys.93.025001}. More details about the daily operation of the 2D ion crystal and the derivation for the Ising model Hamiltonian can be found in our previous work \cite{guo2023siteresolved}.

\section{Alignment of 2D ion crystal with laser wavefront}
Equation~(1) of the main text gives the theoretical Ising coupling coefficients when the laser phase is uniform over the 2D crystal. In practice there can be misalignment between the 2D crystal and the equiphase surface of the laser, e.g. due to the tilting or twisting of the crystal. Then each ion may have a site-dependent initial phase $\varphi_i$, and the theoretical Ising coupling will be modified into $J_{ij} \to J_{ij}\cos(\varphi_i-\varphi_j)$. Before the experiment, we adjust the laser wavefront and the 2D ion crystal to minimize such a site-dependent initial phase.

We calibrate the site-dependent $\varphi_i$ by initializing all the ions in $|S_{1/2},F=0,m_F=0\rangle$, applying a $\pi/2$ pulse using one $411\,$nm laser, and then applying another $\pi/2$ pulse using the other $411\,$nm laser with a random phase. The whole process can be understood as follows.
Without loss of generality, we can use the first laser pulse to define the $\sigma_x$ direction of the optical qubits between $|g\rangle\equiv|S_{1/2},F=0,m_F=0\rangle$ and $|e\rangle\equiv|D_{5/2},F=2,m_F=0\rangle$. Therefore the first $\pi/2$ pulse prepares all the ions into $(|g\rangle+|e\rangle)/\sqrt{2}$. Then the second laser pulse will have a phase shift of $\varphi_i$ in the frame of individual ions, together with a global random phase of $\Delta \phi$ which we add purposely. This leads us to the final state $\{[1+e^{i(\varphi_i+\Delta\phi)}]|g\rangle+[1-e^{-i(\varphi_i+\Delta\phi)}]|e\rangle\}/2$.

From the above final product state, we can compute the expectation values
\begin{equation}
\langle \sigma_z^i\rangle = \cos(\varphi_i+\Delta\phi),
\end{equation}
and
\begin{equation}
\langle \sigma_z^i \sigma_z^j\rangle = \cos(\varphi_i+\Delta\phi)\cos(\varphi_j+\Delta\phi) = \frac{1}{2} \left[\cos(\varphi_i-\varphi_j) + \cos (\varphi_i+\varphi_j+2\Delta\phi)\right].
\end{equation}
If we further average over the random phase $\Delta\phi$, we have
\begin{equation}
\langle \sigma_z^i \sigma_z^j\rangle-\langle \sigma_z^i\rangle\langle \sigma_z^j\rangle = \frac{1}{2} \cos(\varphi_i-\varphi_j),
\end{equation}
which can indicate whether the laser phase is uniform over the 2D crystal.

\begin{figure}
    \centering
    \includegraphics[width=0.6\linewidth]{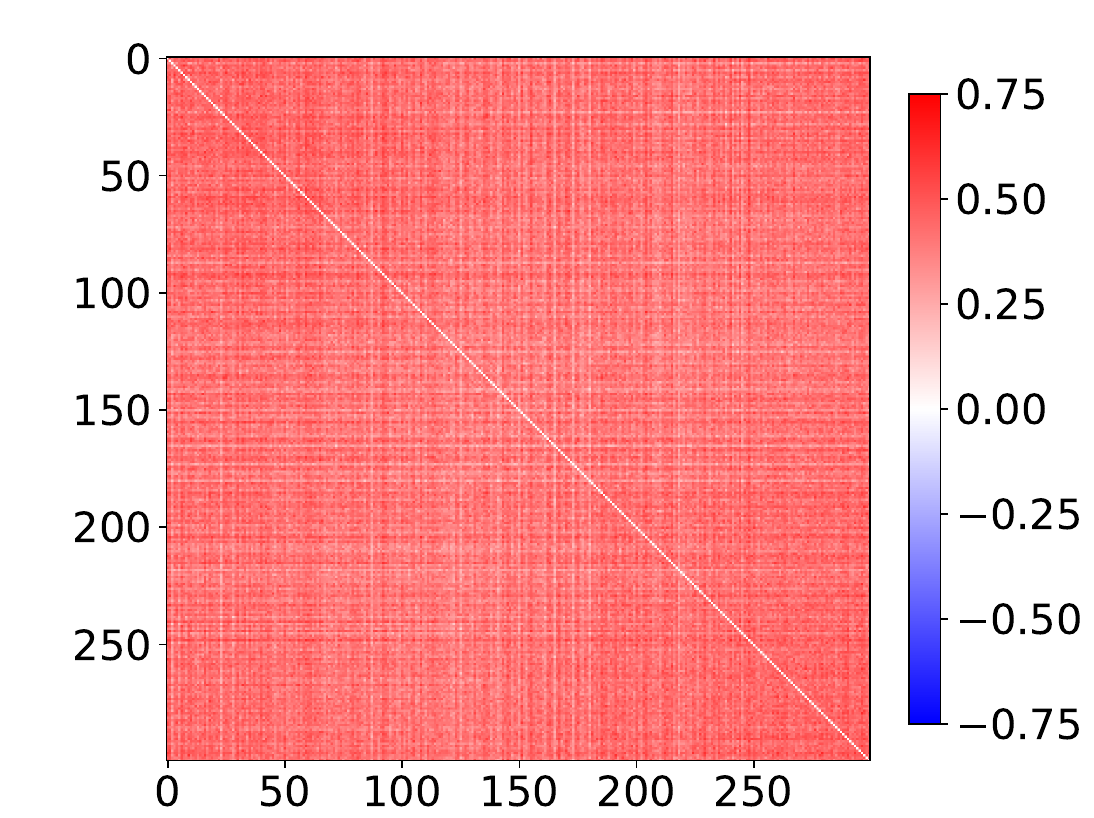}
    \caption{\label{fig:phase} Typical spin-spin correlation under the optimized alignment of the 2D crystal with the laser wavefront.}
\end{figure}

Once the above correlation matrix is measured for the $N=300$ ions, we have two ways to minimize the misalignment: one is to rotate the ion crystal by electric fields, and the other is to rotate the laser beams. Since we want to keep the ion crystal to be micromotion-free along the transverse $y$ direction, which already defines a plane at $y=0$, here we prefer to rotate the laser beams to suppress the spatial oscillation of the spin-spin correlation as much as possible. For the remaining phase fluctuation due to, e.g., the curvature of the laser wavefront, we use the voltages on the $4\times7=28$ DC segments together with an overall DC bias on the RF electrodes to fine-tune the shape of the 2D crystal. Typical measurement results for the final correlation matrix of $N=300$ ions is shown in Fig.~\ref{fig:phase} which indicates a nearly uniform optical phase over the 2D crystal.

\section{Detection of change in the crystal configuration}
As described in Ref.~\cite{guo2023siteresolved}, the configuration of the 2D ion crystal of $N=300$ ions has a typical lifetime of a few minutes under the Doppler cooling laser. However, during the experimental sequence we need to turn off the cooling laser for the desired Hamiltonian evolution. Then the ion configuration has a non-negligible probability to change during the evolution time up to $9\,$ms. In this experiment, although the global laser and the pattern of the few highest phonon modes are not sensitive to the small change in the ion positions, it does influence the site-resolved state detection of individual qubits. Therefore, we add a step to check the crystal configuration during the repetition of the experimental sequences. Specifically, we collect the fluorescence from individual ions during the Doppler cooling stage for each experimental trial using the CMOS camera. Ideally, all the ions will locate at the precalibrated positions and appear bright on the images. However, if the crystal configuration is changed, some of these preselected regions will become empty and be detected as a dark ion.

Note that under our experimental conditions the probability for the configuration to change in each trial is still low. Therefore, once the configuration is changed, it will persist for several rounds of experimental sequences. This allows us to distinguish the configuration change from the occasional SPAM error on random ions.
Specifically, we use the following empirical criteria to identify the configuration change.

\begin{algorithm}[H]
\begin{algorithmic}[1]
\If{In 100 consecutive trials, any one of all the ions is dark for more than 5 times}
    \If{Such problematic ions are bright in any 5 consecutive trials}
    \State{Configuration is not changed in these consecutive trials and their data are kept.}
    \Else
    \State{Configuration is changed. All the data are discarded.}
    \EndIf
\Else
    \If{Any one of all the ions is dark for 3 consecutive trials}
    \State{Configuration is changed in these consecutive trials and their data are discarded.}
    \Else
    \State{Configuration is not changed. All the data are kept.}
    \EndIf
\EndIf
\end{algorithmic}
\end{algorithm}

Also note that, the above criteria can discard not only the change in the ion configuration, but also the occasional leakage to some metastable levels or the formation of e.g. $\mathrm{YbH}^+$ ions. In the experiment, we typically discard about $20\%$ of the data, which depends on the conditions of the experimental setup.

\section{Analytical formulae for $k$-body spin correlations}
We consider a general Ising model Hamiltonian with a longitudinal field
\begin{equation}
H = H_0+H^\prime=\sum_{i<j} J_{ij} \sigma_z^i \sigma_z^i + \sum_i h_i \sigma_z^i \label{eq:Hamiltonian}.
\end{equation}
Following our experimental sequence, we initialize all the spins in $|+\rangle$, evolve the system under the Hamiltonian $H$ for time $t$, and finally measure the individual spins in the $\sigma_x$ basis.

First we consider the single-spin observables $\langle \sigma_x^i \rangle$. We have
\begin{align}
\langle \sigma_x^i(t) \rangle =& \langle +\cdots +|e^{i H t} \sigma_x^i e^{-i H t}|+\cdots +\rangle \nonumber\\
=& \frac{1}{2^n}\sum_{\{s\},\{s^\prime\}} \langle \{s\}|e^{i H t} \sigma_x^i e^{-i H t}|\{s^\prime\}\rangle \nonumber\\
=& \frac{1}{2^n}\sum_{\{s\},\{s^\prime\}} \langle \{s\}|e^{i E(\{s\}) t} \sigma_x^i e^{-i E(\{s^\prime\}) t}|\{s^\prime\}\rangle \nonumber\\
=& \frac{1}{2^n}\sum_{\{s\}-s_i,s_i} \langle \{s\}-s_i|\langle s_i|e^{i E(\{s\}-s_i,s_i) t} \sigma_x^i e^{-i E(\{s\}-s_i,-s_i) t}|\{s\}-s_i\rangle |-s_i\rangle,
\end{align}
where $\{s\}$ represents the set of all spins in the $\sigma_z$ basis, $\{s\}-s_i$ means the set of all but the $i$-th spin, and we use the fact that $\sigma_x^i$ flips the $i$-th spin without affecting other spins.

In the above derivation, we use the function $E$ to represent the energy of a given spin configuration. In general, if we divide the spins into two groups $\{s_A\}$ and $\{s_B\}$, we can express the total energy as
\begin{align}
E(\{s\}) =& E_J(\{s\}) + E_h(\{s\}) \nonumber\\
=& E_J(\{s_A\}) + E_J(\{s_B\}) + E_J(\{s_A\},\{s_B\}) + E_h(\{s_A\}) + E_h(\{s_B\}),
\end{align}
where $E_J$ and $E_h$ represent the energy under the Ising interaction and the longitudinal field, respectively, and $E_J(\{s_A\},\{s_B\})$ is the interaction between the two groups. In addition, we have the symmetry
\begin{align}
E_J(\{s\}) =& E_J(-\{s\}), \nonumber\\
E_h(\{s\}) =& -E_h(-\{s\}), \nonumber\\
E_J(\{s_A\},\{s_B\}) =& -E_J(\{s_A\},-\{s_B\}) = -E_J(-\{s_A\},\{s_B\}).
\end{align}

Using these relations, we get
\begin{align}
\langle \sigma_x^i(t) \rangle =& \frac{1}{2^n}\sum_{\{s\}-s_i,s_i} e^{i E(\{s\}-s_i,s_i) t} e^{-i E(\{s\}-s_i,-s_i) t} \nonumber\\
=& \frac{1}{2^n}\sum_{\{s\}-s_i,s_i} e^{2i [E_J(\{s\}-s_i,s_i) + E_h(s_i)]t} \nonumber\\
=& \frac{1}{2^n}\sum_{\{s\}} e^{2i (\sum_{k\ne i} J_{ki}s_k s_i + h_i s_i )t} \nonumber\\
=& \frac{1}{2}\sum_{s_i}  e^{2i h_i s_i t} \prod_{k\ne i}\cos (2J_{ki}s_i t) \nonumber\\
=& \cos (2h_i t) \prod_{k\ne i}\cos (2J_{ki} t). \label{eq:mag}
\end{align}

Similarly, for the correlation between two spins $i$ and $j$ we have
\begin{align}
\langle \sigma_x^i(t)\sigma_x^j(t) \rangle =& \frac{1}{2^n}\sum_{\{s\}-s_i-s_j,s_i,s_j} e^{i E(\{s\}-s_i-s_j,s_i,s_j) t} e^{-i E(\{s\}-s_i-s_j,-s_i,-s_j) t} \nonumber\\
=& \frac{1}{2^n}\sum_{\{s\}-s_i-s_j,s_i,s_j} e^{2i [E_J(\{s\}-s_i-s_j,s_i) + E_J(\{s\}-s_i-s_j,s_j) + E_h(s_i) +E_h(s_j)]t} \nonumber\\
=& \frac{1}{2^n}\sum_{\{s\}} e^{2i (\sum_{k\ne i,j} J_{ki}s_k s_i + \sum_{k\ne i,j} J_{kj}s_k s_j + h_i s_i + h_j s_j)t} \nonumber\\
=& \frac{1}{4}\sum_{s_i,s_j}  e^{2i (h_i s_i + h_j s_j) t} \prod_{k\ne i,j}\cos [2(J_{ki}s_i + J_{kj}s_j) t] \nonumber\\
=& \frac{1}{2}\cos [2(h_i + h_j) t] \prod_{k\ne i,j}\cos [2(J_{ki} + J_{kj}) t] \nonumber\\
& + \frac{1}{2}\cos [2(h_i - h_j) t] \prod_{k\ne i,j}\cos [2(J_{ki} - J_{kj}) t]. \label{eq:cor}
\end{align}

In principle, by fitting the $N$ single-spin magnetizations and the $N(N-1)/2$ two-spin correlations, we can get all the required parameters. In particular, if we focus on the early-time dynamics, we have
\begin{equation}
\langle \sigma_x^i(t) \rangle \approx 1 - 2 \left(h_i^2 + \sum_{k\ne i} J_{ki}^2\right) t^2 , \label{eq:mag_approx}
\end{equation}
and
\begin{equation}
\langle \sigma_x^i(t)\sigma_x^j(t) \rangle \approx 1 - 2 \left[h_i^2 + h_j^2 + \sum_{k\ne i,j} (J_{ki}^2 + J_{kj}^2)\right] t^2. \label{eq:cor_approx}
\end{equation}
Therefore we have
\begin{equation}
\langle \sigma_x^i(t)\sigma_x^j(t) \rangle - \langle \sigma_x^i(t) \rangle\langle \sigma_x^j(t) \rangle \approx 4 J_{ij}^2 t^2, \label{eq:two_body}
\end{equation}
which already gives the magnitude of each desired $J_{ij}$. However, in practice such early-time dynamics will be sensitive to the SPAM error, so we use the long-time evolution and the analytical formulae to fit the Ising coupling coefficients as described in the main text.

We can also generalize the above formulae to $k$-body correlations
\begin{align}
\langle \sigma_x^{i_1}(t)\cdots\sigma_x^{i_k}(t) \rangle
=& \frac{1}{2^n}\sum_{\{s\}} e^{2i (\sum_l^\prime\sum_k J_{l i_k}s_l s_{i_k} + \sum_k h_{i_k} s_{i_k})t} \nonumber\\
=& \frac{1}{2^k}\sum_{s_{i_1},\cdots,s_{i_k}}  e^{2i (\sum_k h_{i_k} s_{i_k}) t} {\prod_{l}}^\prime \cos \left[2\left(\sum_k J_{l i_k} s_{i_k}\right) t\right] \nonumber\\
=& \frac{1}{2^k}\sum_{s_{i_1},\cdots,s_{i_k}}  \cos \left[2 \left(\sum_k h_{i_k} s_{i_k}\right) t\right] {\prod_{l}}^\prime \cos \left[2\left(\sum_k J_{l i_k} s_{i_k}\right) t\right], \label{eq:k-body}
\end{align}
where $^{\prime}$ in the summation or production represents the set $\{s\}-s_{i_1}-\cdots-s_{i_k}$. The number of terms scales exponentially with $k$, so we do not use all of them for the Hamiltonian learning process, but only use a few randomly chosen sets with $k=3,4,5$ for the validation of our learning results.

In the above derivations we assume ideal evolution under the Hamiltonian of Eq.~(\ref{eq:Hamiltonian}), and we attribute the decay in the magnetization and the correlation to the dynamics under $h_i$'s and $J_{ij}$'s. In practice, however, various noise sources can lead to decoherence in the experiment. For example, slow drifts in the laser intensity or frequency can lead to shot-to-shot fluctuation in these coefficients, which further translates into a Gaussian decay in the measured magnetizations and correlations. Also, as we show in Ref.~\cite{guo2023siteresolved}, the off-resonant phonon excitation can also be regarded as a spin dephasing term when we trace out the phonon modes. Clearly the actual error model can be very complicated depending on the contribution of different sources, and will generally vary with different sites. Here we take a simplified model with $2N$ parameters.

Specifically, for each ion $i$ we assign two parameters $\gamma_{\mathrm{cor}}^i$ and $\gamma_{\mathrm{indep}}^i$ for the correlated and independent decoherence rates, respectively. The correlated decoherence is motivated by the slow drift in the longitudinal fields caused by the global laser intensity. Although they average into a Gaussian decay over different experimental trials through $\langle \cos(h t)\rangle=e^{-(\gamma t)^2/2}$ when $h\sim N(0,\gamma^2)$, still the coherence between different ions is maintained when we consider spin-spin correlations. In contrast, the independent decoherence captures the other error sources that do not maintain the phase coherence among ions. With such terms included, and with the longitudinal field set to zero, the theoretical dynamics for single-spin and multi-spin observables become
\begin{align}
\langle \sigma_x^i(t) \rangle
=& \exp\left\{-[(\gamma_{\mathrm{indep}}^i)^2 + (\gamma_{\mathrm{cor}}^i)^2] t^2\right\} \cdot \prod_{k\ne i}\cos (2J_{ki} t), \\
\langle \sigma_x^i(t)\sigma_x^j(t) \rangle
=& \frac{1}{2}\exp\left\{-[(\gamma_{\mathrm{indep}}^i)^2 + (\gamma_{\mathrm{indep}}^j)^2 + (\gamma_{\mathrm{cor}}^i + \gamma_{\mathrm{cor}}^j)^2] t^2\right\} \nonumber\\
&\quad\quad\times \prod_{k\ne i,j}\cos [2(J_{ki} + J_{kj}) t] \nonumber\\
& + \frac{1}{2}\exp\left\{-[(\gamma_{\mathrm{indep}}^i)^2 + (\gamma_{\mathrm{indep}}^j)^2 + (\gamma_{\mathrm{cor}}^i - \gamma_{\mathrm{cor}}^j)^2] t^2\right\} \nonumber\\
&\quad\quad\times \prod_{k\ne i,j}\cos [2(J_{ki} - J_{kj}) t], \\
\langle \sigma_x^{i_1}(t)\cdots\sigma_x^{i_k}(t) \rangle
=& \frac{1}{2^k}\sum_{s_{i_1},\cdots,s_{i_k}}  \exp\left\{-\left[\sum_k (\gamma_{\mathrm{indep}}^{i_k})^2 + \left(\sum_k \gamma_{\mathrm{cor}}^{i_k} s_{i_k}\right)^2 \right]t^2\right\} \nonumber\\
&\quad\quad\times {\prod_{l}}^\prime \cos \left[2\left(\sum_k J_{l i_k} s_{i_k}\right) t\right].
\end{align}

\begin{figure}
    \centering
    \includegraphics[width=\linewidth]{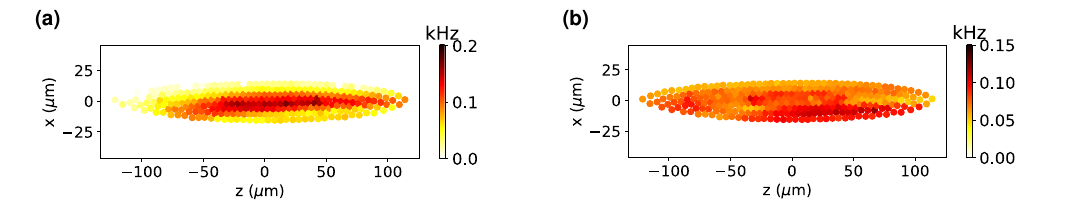}
    \caption{\label{fig:decoherence} Typical fitted decoherence rates of individual ions. (a) $\gamma_{\mathrm{indep}}^i$. (b) $\gamma_{\mathrm{cor}}^i$.}
\end{figure}

If we fit these $2N$ parameters together with the Ising coupling coefficients, since their leading-order effects are both the decay in the magnetization and correlation as shown by Eqs.~(\ref{eq:mag_approx}) and (\ref{eq:cor_approx}), it will largely increase the uncertainty in the learning results. Instead, here we first fit all the $\gamma_{\mathrm{cor}}^i$'s and $\gamma_{\mathrm{indep}}^i$'s with fixed $J_{ij}=0$ by setting a large laser detuning of $80\,$kHz above the COM mode. Then we fix these fitted decoherence rates during the rest of the experiment to learn the desired Hamiltonian when the laser detuning is closer to some of the phonon modes. Note that this separated calibration step is consistent with our error model of a slow drift in the laser-induced AC Stark shift (the longitudinal field) and an additional independent dephasing due to noise sources other than the laser. The calibration results for $\gamma_{\mathrm{cor}}^i$'s and $\gamma_{\mathrm{indep}}^i$'s are shown in Fig.~\ref{fig:decoherence}. Together, they give an average decoherence time of about $1/\sqrt{(\gamma_{\mathrm{indep}}^i)^2 + (\gamma_{\mathrm{cor}}^i)^2}\sim 9\,$ms.

\section{Compensating the leakage error to $D_{5/2}$ and $F_{7/2}$ levels}
Under the off-resonant $411\,$nm laser, there is a small probability that the final population will be in the $D_{5/2}$ or $F_{7/2}$ levels, which is a leakage error from the qubit subspace in the $S_{1/2}$ levels. In the experiment we observe a gradual increase in the leakage probability with the evolution time, which mainly comes from the spontaneous emission from the $D_{5/2}$ levels to the $F_{7/2}$ levels. After electron shelving, such leaked population will be detected as dark, that is, the $|0\rangle\equiv |S_{1/2},F=0,m_F=0\rangle$ state. This will cause bias in the measured magnetizations and correlations.

For small leakage probability $\epsilon_L$, we can measure the leakage rate experimentally and correct it to the first order during the experimental sequence. The idea is to divide the experimental trials into two groups. For one group, we use the original experimental sequence in Fig.~1(b) of the main text. For the other group, we insert a $\pi$ pulse to exchange the $|0\rangle\equiv |S_{1/2},F=0,m_F=0\rangle$ and $|1\rangle\equiv |S_{1/2},F=1,m_F=0\rangle$ states before the measurement. Suppose ideally the $i$-th qubit has probability $p_0^i$ to be in $|0\rangle$ and probability $p_1^i$ to be in $|1\rangle$ ($p_0^i+p_1^i=1$), and suppose now there is probability $\epsilon_L^i$ of the leakage error which may vary for different ions. Now if we evaluate $\langle \sigma_z^i\rangle$ for the two groups of data sets, we will get $(p_0^i-p_1^i)(1-\epsilon_L^i)+\epsilon_L^i$ and $(p_1^i-p_0^i)(1-\epsilon_L^i)+\epsilon_L^i$, respectively, without or with the $\pi$ pulse. From their average, we obtain the leakage probability $\epsilon_L^i$, while their difference gives $2(p_0^i-p_1^i)(1-\epsilon_L^i)$, which is proportional to the ideal single-spin magnetization $p_0^i-p_1^i$ and can be recovered by dividing $(1-\epsilon_L^i)$. Following similar derivations, we can use this method to correct the $k$-body spin correlations $\langle \sigma_z^{i_1}(t)\cdots\sigma_z^{i_k}(t) \rangle$ to the first order by dividing $\prod_{l=1}^k (1-\epsilon_L^{i_l})$.

\begin{figure}
    \centering
    \includegraphics[width=0.6\linewidth]{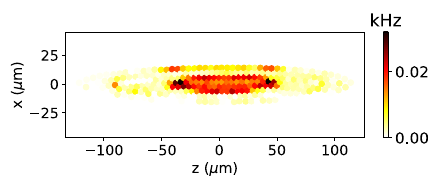}
    \caption{\label{fig:leakage} Typical leakage rates of individual ions.}
\end{figure}

For each ion, we can further perform a linear fit for the leakage probability versus time and obtain the leakage rates shown in Fig.~\ref{fig:leakage}. The central ions feel higher $411\,$nm laser intensity and typically have higher leakage rates, with a typical timescale above $30\,$ms, much longer than our evolution time.

\section{Theoretical scaling of the precision}
As described in the main text, we use the relative energy difference $\epsilon(J^{(1)},J^{(2)})\equiv \langle |E(J^{(1)})-E(J^{(2)})|\rangle/\sqrt{\delta E(J^{(1)}) \cdot \delta E(J^{(2)})}$ between the learning results $J_{ij}^{(1)}$'s and $J_{ij}^{(2)}$'s on independent data sets to characterize the learning precision. Specifically, for any given spin configuration $\{s\}$, we can evaluate the energy for the two Hamiltonians $E(\{s\};J^{(1)})$ and $E(\{s\};J^{(2)})$. Then we compute the numerator as the average of $|E(\{s\};J^{(1)})-E(\{s\};J^{(2)})|$ over the randomly sampled spin configurations, and we define $\delta E(J^{(1)})$ and $\delta E(J^{(2)})$ as the standard deviations of $E(\{s\};J^{(1)})$ and $E(\{s\};J^{(2)})$ over random spin configurations.

Next we analyze the scaling of the precision $\epsilon$ versus the system size $N$. When fitting the $O(N^2)$ parameters of $J_{ij}$'s, the different parameters are correlated through their covariance matrix, so we expect their fluctuation to be on the same order which we denote as $\delta J_{ij}\sim\delta$. We have $\delta\propto 1/\sqrt{M}$ depending on the sample size $M$. Now if we compute the energy difference due to the $\delta J_{ij}$ terms on random spin configurations, we are basically performing a random walk with $O(N^2)$ steps. Therefore the numerator can be estimated to be $O(N\delta)$. On the other hand, for the denominator we consider two cases: (i) We couple dominantly to a single phonon mode such that we have an all-to-all coupling $J_{ij}\sim O(J_0)$; (ii) We couple to all the phonon modes with roughly a power-law decay $J_{ij}\sim J_0/\|\vec{r}_i-\vec{r}_j\|^\alpha$ \cite{RevModPhys.93.025001}. In both cases, to observe nontrivial dynamics in the magnetizations and correlations, we want $\sum_{k\ne i} J_{ik}^2 T^2 \sim O(1)$ according to Eqs.~(\ref{eq:mag_approx}) and (\ref{eq:cor_approx}) where $T$ is the total evolution time. In the first case it means $J_0\sim 1/\sqrt{N} T$ and in the second case it gives $J_0\sim 1/T$. In both cases, after averaging over random spin configurations, we find the standard deviation of the total energy to be $O(\sqrt{N}/T)$, thus the precision will scale as $\sqrt{N}\delta T$.

On the other hand, if the ground state properties are desired, then in the denominator we should not compare with the fluctuation of the energy but with the ground state energy itself. Suppose there to be no significant frustration, we expect the ground state energy to scale at least as $O(N)$, then the precision will be $O(1)$, no longer degrading with the system size.

\section{Fitting anharmonic trap potential}
To compute the Ising coupling coefficients theoretically using Eq.~(1) of the main text, we need to know the accurate phonon mode structure of the ion crystal, which is in principle a classical problem and can be solved given the external trap potential. The trap potential is in principle governed by the design of the trap electrodes and the voltages we apply on them. However, due to the fabrication errors and various experimental noises, the theoretically computed trap potential can often deviate from the actual one. Therefore we choose to fit the trap potential from the available information, including the measured ion positions in the 2D crystal at the precision of about $1\,\mu$m, the frequencies of a few resolvable phonon modes at the precision of about $1\,$kHz, and the excitation pattern of these modes. As the phonon spectrum becomes denser at the low-frequency side, here we only use the frequencies of the 10 highest modes and one lowest frequency. Also we can apply a weak global $411\,$nm laser pulse on the blue sideband of a resolved mode $k$, then the excitation probability of each ion $i$ will be proportional to $b_{ik}^2$.

Often, a harmonic trap is still a reasonable approximation to the trap potential, and here we add a few anharmonic terms as small perturbation.
In particular, we assume that the potential can be described by some low-order polynomials which vary slowly in space. Otherwise, if the potential is fast oscillating from ion to ion, then there will be too many parameters to fit from the available information.

Even if we truncate to quartic polynomials, the potential already contains a large number of parameters and a straightforward fitting will be both inefficient and likely to be trapped to unphysical local minima. Instead, we divide the fitting procedure into five stages as shown below:

\begin{center}
\begin{tabular}{c|c|c|c}
\hline
fitting stage & fitting operations & fitted terms & objective functions\\
\hline
1 & initial quadratics & $x^2$, $z^2$ & positions, frequencies\\
2 & cubics within $xz$ plane& $x^2$, $z^2$, $x^3$, $x^2z$, $xz^2$, $z^3$ &positions\\
3 & refine quadratics & $x^2$, $z^2$ &positions, frequencies\\
4 & symmetric quartics & $x^2$, $z^2$, $y^2z^2$, $y^2x^2$, $x^2z^2$, $z^4$ &positions, frequencies\\
5 & cubics along $y$& $xy^2$, $zy^2$ &mode vectors\\
\hline
\end{tabular}
\end{center}

In the first stage, we give an initial fitting of the harmonic terms $x^2$ and $z^2$ within the plane of the 2D ion crystal. The $y^2$ term is measured as the single-ion trap frequency along the perpendicular $y$ axis through the resolved sideband transition with high precision, and is fixed during the whole fitting procedure. For any given coefficients for the $x^2$ and $z^2$ terms, we use the measured ion positions as the starting point to search the equilibrium positions under their Coulomb interaction and further solve the collective phonon mode frequencies. Then we compare them with the measured values and further improve the fitting results by minimizing this cost function.

In the second stage, we add cubic terms to better fit the ion positions inside the $xz$ plane. Because the equilibrium positions of all the ions have $y=0$, we can drop all the terms containing $y$ and only consider $x^3$, $x^2z$, $xz^2$ and $z^3$ terms.

However, the cubic terms are not sufficient to give a good fitting to the ion positions and mode frequencies, so we further consider the quartic terms. To have a good starting point to fit the quartic terms as perturbations, we insert the third stage to refine the quadratic terms with the previous cubic terms fixed. Then in the fourth stage we add some quartic terms. Due to the reflection symmetry of the designed electrodes, here we only consider symmetric terms like $y^2z^2$, $y^2x^2$, $x^2z^2$ and $z^4$, and we expect the other asymmetric terms to be subdominant. In fact, the asymmetry of the ion crystal can already be captured by the cubic terms above. Also here we drop the $x^4$ term because the size the 2D crystal is much smaller in the $x$ direction than that in the $z$ direction.

In the last stage, we further fit cubic terms along the $y$ direction through their influence to the transverse phonon modes. Here we focus on $x y^2$ and $z y^2$ terms, which can be regarded as a site-dependent trap frequency along the $y$ direction. They will not change the equilibrium positions of the ions, but will affect the phonon modes. On the other hand, terms like $x^2y$, $z^2y$ and $xyz$ correspond to a site-dependent force perpendicular to the crystal. Their main effect is a twist in the crystal that cannot be measured from the image of the ions. Also, after the small twist of the crystal, these forces are already cancelled by the harmonic terms as well as the Coulomb interaction between ions, so we expect their influence on the phonon modes to be higher order and hence do not include them in this fitting. Finally, the $y^3$ does not affect the equilibrium positions or the transverse phonon modes and therefore is not considered either.

In the above fitting, we do not consider the micromotion of the ions because in our case the largest micromotion amplitude is still smaller than the inter-ion distance. In the future, we may also take the micromotion into consideration to get a more accurate description of the phonon modes for larger ion crystals \cite{PhysRevA.103.022419,Landa2012,Landa2012_2}.

%